\documentclass[12pt]{article}
\usepackage{amsmath}
\usepackage{amsfonts}
\usepackage{amssymb}
\usepackage{graphicx}
\begin{document}
\setcounter{page}{0} \topmargin 0pt
\renewcommand{\thefootnote}{\arabic{footnote}}
\newpage
\setcounter{page}{0}

%\begin{titlepage}

\begin{center}
{\Large {\bf The Legacy of Ken Wilson\footnote{Talk given at the Ken Wilson Memorial Session, StatPhys 25, Seoul, July 2013}}}\\

\vspace{2cm}
{\large John Cardy$^{a,b}$\\}  \vspace{0.5cm} {\em $^{a}$Rudolf
Peierls Centre for Theoretical Physics\\ 1 Keble Road, Oxford OX1
3NP, UK\\}  \vspace{0.2cm} {\em $^{b}$All Souls College, Oxford\\}

\vspace{2cm}

\end{center}

%\vspace{1cm}

%\begin{abstract}
%\end{abstract}

%\end{titlepage}

I should begin this short presentation with some apologies. First, I was not a close colleague of Ken Wilson, and indeed, apart from attending some of his seminars and lectures, met him on only a couple of occasions. There are people in this audience who were much better acquainted with him. When he was making his major contributions I was a graduate student, then a postdoc, in high energy physics, and was not immediately aware of his work. In 1974, however, his article with John Kogut `The Renormalization Group and The $\epsilon$- expansion'
appeared in Physics Reports \cite{WK}, and this became my bible, along with that of many others I suspect. My sole qualification for giving this talk is that I can claim to represent a generation of theoretical physicists for which Wilson's ideas became a central pillar of our research. In an accompanying paper, Leo Kadanoff gives a much more authoritative view of a  contemporary \cite{Kad}.

The second apology is that in fifteen minutes I can barely do this subject justice. I very much hope that in due course a more detailed collection of memorial articles will appear. 

I have chosen to structure this talk in a `Before Wilson/After Wilson' format. That is, I will try to compare what was known and understood before, and after, Wilson was active. Of course, he was not the only person making important contributions at this time, but this is neither the place, nor do I have the historical knowledge, to give a detailed account of exactly who did what, and when, in what was a rapidly developing field. What is clear is that Wilson's role, spread across a wide swath of theoretical physics, was central and unique. 

First, however, an abbreviated CV. Kenneth Geddes Wilson was born in 1936 in Waltham, Mass., eldest child of E.~Bright Wilson, a prominent quantum chemist at Harvard. Ken was himself an undergraduate there. In addition to his studies he was an athletics star, and this attraction to physical exercise, and to the outdoors, seems to have remained with him all his life. He did his graduate studies at CalTech under the supervision of Murray Gell-Mann, but this was the time when many other reknowned theoretical physicists, such as Richard Feynman, were also around. Wilson was appointed to the faculty at Cornell in 1963, where he stayed until 1988 and did most of the work in theoretical physics for which he is known. After this he moved
to Ohio State, from which he retired in 2008. Among his many honours was the award of the inaugural Boltzmann Medal in 1975, the Wolf Prize in 1980 (which he shared with Michael Fisher and Leo Kadanoff), and of course the Nobel Prize for Physics in 1982, ``for his theory for critical phenomena in connection with phase transitions''.

So now for the Before/After list:

$\bullet$ \em Before Wilson, quantum field theory and statistical mechanics were separate subjects.\em

Afterwards, it became accepted that they are different sides of the same coin, and that concepts and methods applicable to one may be freely be transferred to the other. For example, in a very well-defined sense, the Ising model \em is \em $\lambda\phi^4$ field theory. As a graduate student at Cambridge I attended courses in quantum field theory, where we studied Feynman diagrams and renormalisation, and in statistical mechanics, where we studied, among other things, Onsager's solution. I was amazed to discover, a few years later, that these subjects are closely related. Of course, nowadays this is accepted as commonplace, and indeed the topics of Feynman diagrams, renormalisation and critical behaviour are often taught as part of the same introductory graduate course in field theory, as applied to both high energy and condensed matter physics. It was Wilson who also realised that gauge theories may also be realised as a kind of statistical physics, with degrees of freedom living on the edges, rather than the vertices of a lattice. This led to a further strengthening of the ties between the two disciplines.

But at the time, Wilson's ideas were met with skepticism. A contemporary \cite{Brezin} describes his 1971 lectures in Princeton thus:

\em ``Ken was not easy to understand : not only he had his
idiosyncratic way  of dealing with theories as objects that lead to explicit
practical calculations rather than abstract concepts, but he was forcing us to
look at the most beautiful theories as merely ``effective'' or ``emergent'' rather
than fundamental. This was met initially  with enormous skepticism...''\em
\footnote{As an aside, I find it interesting that these comments might equally well have been made of another brilliant Nobel prize-winning physicist: R.~Feynman, who, although differing totally in personality from Wilson, had a similar philosophy of `shut up and calculate'.}

$\bullet$ \em Before Wilson, the renormalisation group was a set of half-formed ideas, both in field theory and in statistical physics.\em

As mentioned above, many people contributed to the development of the RG. In field theory it had already been observed (and, perhaps erroneously, named) as a property of renormalised field theory - yet there it seemed to be almost a tautology.
In critical phenomena, the idea of block spin transformations helped give a framework for scaling. But it seemed to produce no quantitative results until Wilson came along. He realised that the RG really acts in an infinite-dimensional space of all possible hamiltonians. Despite this complexity, it could be forged into a useful computational method. The central idea of fixed points and relevant/irrelevant couplings cut this down to a manageable problem. In some cases, for example the $\epsilon$-expansion, a small parameter made calculations analytically tractable, although this was not a necessary feature of the program. In order to do this, Wilson independently realised that Feynman diagrams can be computed equally well in non-integer dimensions. Later workers related much of this to ideas in renormalised quantum field theory, but Wilson's approach is still more general. 

$\bullet$ \em Before Wilson, the subject of critical phenomena was a mass of poorly understood experimental data and a few exact results. \em

It was realised, for example, that the critical exponents of Ising-like magnets and liquid-gas critical points appeared to be the same (an example of universality). It was understood that mean field theory often gives incorrect predictions for these exponents, but attempts to take into account the effects of fluctuations were uncontrolled. Onsager's exact solution of a logarithmic anomaly in the specific heat of the 2d Ising model was incorrectly applied to other systems. A whole Greek alphabet of critical exponents were defined, and scaling relations between them (some rigorously proved as bounds) were postulated.

The development of the RG gave an immediate explanation of universality in terms of flows in the vicinity of a fixed point.
This simple observation has the power to relate a mass of observations on widely differing physical systems. The scaling relations were shown to follow as exact relations, and the Greek alphabet was reduced to a small number of universal numbers, the RG eigenvalues. 

At the same time, the idea of irrelevant operators predicted the form of corrections to scaling, permitting sensible fitting of data beyond naive log-log plots. The RG explanation of cross-over behaviour between fixed points gave both a qualitative and quantitative description of systems with weak symmetry-breaking interactions, a feature of most real materials. The RG also gave a simple framework for understanding finite-size effects, essential for interpreting numerical simulations, an increasingly important tool as computers became more powerful.

$\bullet$ \em Before Wilson, the role of renormalisation in field theory was itself poorly understood.\em

Although it had of course enjoyed great success with the accurate predictions of QED, the manipulation of infinities was still viewed with suspicion. And the fact that perturbative quantum field theory seemed unable to account for the strong and weak interactions seemed to make this a one-off success. 

But, after Wilson, it came to be realised that the RG is perhaps the best way to understand renormalisation itself in field theory. Moreover, the ideas described in the quotation above, that all field theories are merely effective descriptions of nature at particular energy or length scales, began to take root. The requirement of renormalisability came to be seen merely as a simple way of ensuring that predictions on one scale were effectively decoupled from the detailed description of other scales. In this light, even non-renormalisable theories became acceptable as long as they were not used outside their domain of applicability. The idea of \em emergence\em, central to modern scientific thinking, is encapsulated in the RG.
Nowadays we accept that the field theories used in particle physics are no more `fundamental' than those used in describing condensed matter systems or the behaviour of bacteria in colonies. 

The RG, often renamed as `multi-scale analysis' has also come to be used as a tool in the rigorous treatment of critical behaviour, and in constructive field theory. It has also been very successfully, and rigorously, applied in describing the transition to chaos in dynamical systems.

Another of Wilson's important ideas, the operator product expansion (OPE), should also be mentioned in this context. He was the first to realise that it should hold in a theory (the Thirring model) with non-trivial scaling exponents, not just as a property of free field theory. This paved the way for conformal field theory, as developed by Polyakov and 
others\footnote{Interestingly, Wilson seems to have believed that the  OPE and the RG are unconnected \cite{interview}.}, and also the analysis of deep inelastic electron scattering experiments, essential to reveal the physical existence of quarks.

$\bullet$ \em Before Wilson, colour confinement in gauge theories was only a hypothesis.\em

Even after the earlier ideas of the quark model were incorporated into an SU$(3)$ non-Abelian gauge theory, and this was shown to have the essential RG property of asymptotic freedom at short distances necessary to explain deep inelastic and other high energy experiments, it still had to be assumed that these theories somehow led to confinement of quarks and gluons. By realising gauge theories on a lattice, in a formalism in which the gauge group appears rather than just its Lie algebra, the property of confinement at strong coupling became clear from the outset. Wilson's formulation of lattice gauge theory (supplemented by later prescriptions for dealing with the fermion doubling problem) became, once computational power was available, a major tool in particle physics. As in many things, Wilson was well ahead of his time. The idea of the Wilson loop as a new kind of non-local order parameter, originally devised to test for confinement,  is nowadays as commonly used in condensed matter applications as in high energy physics, as the ideas of lattice gauge theory are taken over to describe topological states of matter. 

$\bullet$ \em Before Wilson, the use of computers to understand fundamental physics was suspect.\em

Although computers were by this time already widely used for numerical calculations and simulations, fundamental physics was still about doing analytic calculations or making suitable approximations. Nowadays we use the power of computers to make fundamental progress, in all fields of theoretical physics. We use them to test severely our analytic methods in regimes where real experiments are impossible, and rely on them for the increasing number of problems which appear to be otherwise completely intractable. 

Wilson was fascinated by the latent power of computers, and really taught us how to think in the way that a computer might -- what can we do this way that we cannot otherwise? One of Wilson's major achievements was a numerical solution of the long-standing Kondo problem using numerical RG methods. Asked how he was inspired to do this, he gave the following very illuminating reply \cite{interview}:

\em Q: And looking at the Kondo effect at that stage, does the stimulation for that come from Phil Anderson?

KW: No. It comes from my utter astonishment at the capabilities of the Hewlett-Packard pocket calculator, the one that does exponents and cosines. And I buy this thing and I can't take my eyes off it and I have to figure out something that I can actually do that would somehow enable me to have fun with this calculator $\ldots$ What happened was that I worked out a very simple version of a very compressed version of the Kondo problem, which I could run on a pocket calculator. And then I realize that this was something I could set up with a serious calculation on a big computer to be quantitatively accurate.\em

Wilson's solution was a turning point in fundamental theoretical physics, and was later adapted by his student Steve White in the density-matrix RG method. This has been the single most successful numerical method for understanding strongly coupled systems, particularly in one dimension, and has also led to a deeper understanding in terms of quantum entanglement, matrix product states and tensor network methods.

Wilson's attempts to apply numerical RG ideas in other contexts were not always successful: for example in later years, and perhaps inspired by his father, he tried to attack quantum chemistry, where unfortunately the separation of length scales characteristic of critical behaviour does not necessarily hold. However at the same time, his early advocacy of the setting up of a national network of super-computer centres was to bear considerable fruit.

He was indeed a computer visionary, in some ways comparable to the founders of major computer platforms:

\em ``In 1976, he gave us a lecture about how, one day, we would be all be sitting on the beach with personal computers that ran UNIX and using them to play games that involved exploration in three dimensions.   
In 1976, that seemed like a dream.  It is amazing how far we have come in that time \cite{peskin}.'' 
\em

$\bullet$ \em Before Wilson, early physics education was not a priority for research physicists.\em

I think it can be fairly said that most research physicists are more concerned that the supply of qualified graduate students does not dry up than at the level of scientific ignorance among the general public, except where it might help justify their level of research funding. However Ken Wilson became much more involved with physics education, particularly at the elementary level, and used his Nobel laureate status as a way of influencing educators and policy makers. He advocated a hands-on approach which he called `physics by inquiry', incorporating the ideas of major educational and sociological thinkers of the time. This led to reforms at both the state and national level.

$\bullet$ To summarise these remarks in one sentence,  I would say that Ken Wilson's advocacy of large scale computing revolutionised our view of theoretical physics,
but his understanding of the renormalisation group revolutionised our view of the whole world.

\em Acknowledgements. \em I warmly thank the following close associates of Ken Wilson for taking the time to respond to my enquiries while I was writing this talk: Edouard Br\'ezin, Michael Fisher, Leo Kadanoff, Michael Peskin, and Steve White.
Any errors of fact or interpretation are the fault entirely of the author.

\end{document}